\documentclass[letterpaper,notoc]{CECS}
\title{Poincar\'{e} invariant gravity with local supersymmetry as a gauge theory for the M-algebra}
\author{Mokhtar Hassa\"{\i}ne, Ricardo Troncoso and Jorge Zanelli\footnote{{\it E-mail:} {\tt  hassaine@cecs.cl, ratron@cecs.cl, jz@cecs.cl}}\\ Centro de Estudios Cient\'{\i}ficos (CECS), Casilla 1469, Valdivia, Chile.  }
\preprint{{\tiny CECS-PHY-03/03} }

\abstract{
Here we consider a gravitational action having local Poincar\'{e} invariance
which is given by the dimensional continuation of the Euler density in ten
dimensions. It is shown that the local supersymmetric extension of this
action requires the algebra to be the maximal extension of the $\mathcal{N}%
=1 $ super-Poincar\'{e} algebra. The resulting action is shown to describe a
gauge theory for the M-algebra, and is not the eleven-dimensional
supergravity theory of Cremmer-Julia-Scherk. The theory admits a class of
vacuum solutions of the form $S^{10-d}\times X_{d+1}$, where $X_{d+1}$ is a
warped product of $\Bbb{R}$ with a $d$-dimensional spacetime. It is shown
that a nontrivial propagator for the graviton exists only for $d=4$ and
positive cosmological constant. Perturbations of the metric around this
solution reproduce linearized General Relativity around four-dimensional de
Sitter spacetime.}

\begin{document}


\section{Introduction}

A consensus has emerged in the high energy community that a consistent
unified theory of all interactions and matter should be formulated in some
dimension higher than four. Strong theoretical evidence, both in
supergravity and in string theory, leads to conjecture the existence of an
underlying fundamental theory in eleven dimensions \cite{CJS}, \cite
{Townsend-D, Witten}. This is nowadays called M-Theory (see, $e.g$., \cite
{Schwarz}). The standard procedure to link the higher dimensional theory
with four-dimensional physics has been either to compactify the extra
dimensions by the Kaluza-Klein reduction (see, e.g., \cite{Duff-Nilsson-Pope}%
), or through some more recent alternatives \cite{Alternative}.

In these frameworks, however, the physical spacetime dimension is an input
rather than a prediction of the theory. In fact, in standard theories whose
gravitational sector is described by the Einstein-Hilbert action, there is
no obstruction to perform dimensional reductions to spacetimes of dimensions
$d\neq 4$. Then the question arises, since eleven-dimensional Minkowski
space is a maximally (super)symmetric state, and the theory is well-behaved
around it, why the theory does not select this configuration as the vacuum,
but instead, it chooses a particular compactified space with less symmetry.
An ideal situation, instead, would be that the eleven-dimensional theory
dynamically predicted a low energy regime which could only be a
four-dimensional effective theory. In such a scenario, a background solution
with an effective spacetime dimension $d>4$ should be expected to be a false
vacuum where the propagators for the dynamical fields are ill-defined, lest
a low energy effective theory could exist in dimensions higher than four.

In this paper, a new eleven-dimensional theory sharing some of these
features is constructed. Indeed, for this theory, eleven-dimensional
Minkowski spacetime is a maximally supersymmetric solution that would be a
natural candidate for the vacuum. However, propagators around this
background are ill-defined and hence it is a sort of false vacuum. On the
other hand, the theory admits vacuum geometries of the form $S^{10-d}\times
X_{d+1}$, where $X_{d+1}$ is a domain wall whose worldsheet is a $d$%
-dimensional constant curvature spacetime $M_{d}$.\ These solutions exist
only if $M_{d}$ has a non-negative cosmological constant, and the graviton
can only propagate provided $M_{d}$ is a four-dimensional de-Sitter space.\
Moreover, the gravitational perturbations reproduce linearized General
Relativity in four dimensions. Thus, the resulting four-dimensional
effective theory is indistinguishable from gravity with positive
cosmological constant in perturbation theory. Our motivation to choose
eleven dimensions is to explore new geometrical and dynamical structures
that are expected to exist in $d=11$, and could be regarded as new ``cusps''
of M-theory (see e. g. \cite{WNS}). The theory presented here is not
equivalent to the Cremmer-Julia-Scherk supergravity in eleven dimensions
\cite{CJS}.

The gravitational action we propose is selected by requiring local
Poincar\'{e} invariance and is given by the dimensional continuation of the
Euler density in ten dimensions. Its local supersymmetric extension requires
the algebra to be the maximal extension of the $\mathcal{N}=1$
super-Poincar\'{e} algebra in eleven dimensions, commonly known as the
M-algebra. This algebra is spanned by the set $G_{A}=\{J_{ab},P_{a},Q_{%
\alpha },Z_{ab},Z_{abcde}\}$, where $J_{ab}$ and $P_{a}$ are the generators
of the Poincar\'{e} group and $Q_{\alpha }$ is a Majorana spinor supercharge
with anticommutator \cite{Townsend-L}
\begin{equation}
\{Q_{\alpha },Q_{\beta }\}=\left( C\Gamma ^{a}\right) _{\alpha \beta
}P_{a}+(C\Gamma ^{ab})_{\alpha \beta }Z_{ab}+(C\Gamma ^{abcde})_{\alpha
\beta }Z_{abcde}\;.  \label{MAlgebra}
\end{equation}
The charge conjugation matrix $C$ is antisymmetric, and the ``central
charges'' $Z_{ab}$ and $Z_{abcde}$ are tensors under Lorentz rotations but
otherwise Abelian generators\footnote{%
In standard eleven-dimensional supergravity, these generators correspond to
the ``electric'' and ``magnetic'' charges of the $M2$ and $M5$ branes,
respectively. Note that, contrary to the case in standard supergravity, the
generators of diffeomorphisms ($\mathcal{H}_{\mu }$) are absent from the
right hand side of (\ref{MAlgebra}).}. As shown below, the algebra fixes the
field content to include, apart from the graviton $e_{\mu }^{a}$ , the spin
connection $\omega _{\mu }^{ab}$ and the gravitino $\psi _{\mu }$, two
one-form fields $b_{\mu }^{ab}$, $b_{\mu }^{abcde}$, which are rank two and
five antisymmetric tensors under the Lorentz group, respectively. The local
supersymmetry transformations close off-shell without requiring auxiliary
fields. As will be seen below, the supersymmetric Lagrangian can be
explicitly written as a Chern-Simons form. It is known that for Chern-Simons
theories bosonic and fermionic degrees of freedom do not necessarily match,
since there exists an alternative to the introduction of auxiliary fields
(see e. g. \cite{HIPT}). Indeed, the matching may not occur when the
dynamical fields are assumed to belong to a connection instead of a
multiplet for the supergroup \cite{Tr-Z}.


\section{Gravitational sector}

In dimensions higher than four, under the same assumptions of General
Relativity in four dimensions (i.e., general covariance, second order field
equations for the metric), the so-called Lovelock actions are obtained \cite
{Lovelock}, which include the Einstein-Hilbert lagrangian as a particular
case. In general, these lagrangians are linear combinations of the
dimensional continuations of the Euler densities from all lower dimensions
\cite{Zumino} and therefore contain higher powers of the curvature. Since
the action can be expressed in terms of differential forms without using the
Hodge dual, it is easy to see why these theories do not yield higher
derivative field equations. In the first order formalism (analogous to
Palatini's) the field equations can only involve first order derivatives of
the dynamical fields (for more on this, see \cite{Troncoso-Zanelli}).
Furthermore, if one then imposes the torsion to vanish the field equations
become at most second order. In the vanishing torsion sector, the theory has
the same degrees of freedom as General Relativity \cite{Teitelboim-Zanelli}.

Without imposing the torsion constraint, the field equations remain first
order, even if one couples this theory to other $p$-form fields without
involving the Hodge. In fact, in this way it is impossible to generate
higher derivative terms in this theory.

An action containing (\ref{MAlgebra}) as a local symmetry must be, in
particular, invariant under local translations,
\begin{equation}
\delta e^{a}=D\lambda ^{a}=d\lambda ^{a}+\omega _{\,b}^{a}\lambda
^{b},\delta \omega ^{ab}=0\;.  \label{PoincareTrans}
\end{equation}
The only gravitational action in eleven dimensions constructed out of the
vielbein $e^{a}$ and the spin connection $\omega ^{ab}$, leading to second
order field equations for the metric, invariant under diffeomorphisms and
local Poincar\'{e} transformations is given by \cite
{Chamseddine,Troncoso-Zanelli}
\begin{equation}
I_{G}[e,\omega ]=\int_{M_{11}}\epsilon _{a_{1}\cdots
a_{11}}R^{a_{1}a_{2}}\cdots R^{a_{9}a_{10}}e^{a_{11}}.  \label{I_G}
\end{equation}
Here $R^{ab}=d\omega ^{ab}+\omega _{\;c}^{a}\omega ^{cb}$ is the
curvature two-form, and wedge product between forms is understood \cite
{Footnote0}. For the reason given above, we take $I_{G}$\ as the
gravitational sector of our theory rather than the Einstein-Hilbert action
which, is not invariant under (\ref{PoincareTrans}) \cite{Footnote1}. The
Lagrangian in (\ref{I_G}) is the ten-dimensional Euler density continued to
eleven dimensions and contains the degrees of freedom of eleven dimensional
gravity \cite{Teitelboim-Zanelli}.

A local Poincar\'{e} transformation acting on the dynamical fields is a
gauge transformation $\delta _{\lambda }A=d\lambda +[A,\lambda ]$, with
parameter $\lambda =$ $\lambda ^{a}P_{a}+\frac{1}{2}\lambda ^{ab}J_{ab}$,
provided $e^{a}$ and $\omega ^{ab}$ are the components of a single
connection for the Poincar\'{e} group, $A=e^{a}P_{a}+\frac{1}{2}\omega
^{ab}J_{ab}.$ This observation will be the guiding principle for the
construction of a locally supersymmetric extension of $I_{G}$.


\section{Supersymmetric extension}

A natural way to construct a locally supersymmetric extension of (\ref{I_G})
without breaking local Poincar\'{e} invariance is that the extra fields
required by supersymmetry enter on a similar footing with the original
fields. In other words, all dynamical fields will be assumed to belong to a
connection for a supersymmetric extension of the Poincar\'{e} group. This
approach strongly deviates from the standard assumption in supergravity,
where the fields are assumed to belong to a multiplet. As we shall see now,
the M-algebra emerges naturally from our approach. The simplest tentative
option would be to consider the $\mathcal{N}=1$ super Poincar\'{e} algebra
without central extensions. However, this\ possibility must be ruled out.
Indeed, in this case, the connection would be extended by the addition of a
gravitino as $A\rightarrow A+\psi Q/\sqrt{2}$, and the gauge generator would
change as $\lambda \rightarrow \lambda +\epsilon Q/\sqrt{2}$, where $%
\epsilon $\ is a zero-form Majorana spinor. This would fix the
supersymmetric transformations to be $\delta e^{a}=\bar{\epsilon}\Gamma
^{a}\psi /2$, $\delta \psi =D\epsilon $ and $\delta \omega ^{ab}=0$. Then
the variation of (\ref{I_G}) under supersymmetry can be cancelled by a
kinetic term for the gravitino of the form
\begin{equation}
I_{\psi }=-\frac{1}{6}\int_{M_{11}}R_{abc}\bar{\psi}\Gamma ^{abc}D\psi ,
\label{I-Psi}
\end{equation}
where $R_{abc}:=\epsilon _{abca_{1}\cdots a_{8}}R^{a_{1}a_{2}}\cdots
R^{a_{7}a_{8}}$. However, the variation of $I_{\psi }$ produces, in turn, an
extra piece which cannot be cancelled by a local Lagrangian for $e^{a}$, $%
\omega ^{ab}$, and $\psi $, and hence the super Poincar\'{e} algebra is not
rich enough to ensure the off-shell supersymmetry of the action.
Nevertheless, following the Noether procedure, it can be seen that
supersymmetry can be achieved introducing additional bosonic fields. These
fields can only be either a second-rank or a fifth-rank tensor one-forms $%
b^{ab}$, and $b^{abcde}$, that transform like $\bar{\epsilon}\Gamma
^{ab}\psi $ and $\bar{\epsilon}\Gamma ^{abcde}\psi $, respectively. Assuming
that the dynamical fields belong to a single connection for a supersymmetric
extension of the Poincar\'{e} group, the only option that brings in these
extra bosonic fields is to consider the M-algebra (\ref{MAlgebra}), which
also prescribes their supersymmetry transformations in the expected form.
This means that the field content is given by the components of a single
fundamental field, the M-algebra connection,
\begin{equation}
A=\frac{1}{2}\omega ^{ab}J_{ab}+e^{a}P_{a}+\frac{1}{\sqrt{2}}\psi ^{\alpha
}Q_{\alpha }+b^{ab}Z_{ab}+b^{abcde}Z_{abcde}\;,  \label{M-connection}
\end{equation}
and hence, the required local supersymmetry transformations are obtained
from a gauge transformation of the M-connection (\ref{M-connection}) with
parameter $\lambda =1/\sqrt{2}\epsilon ^{\alpha }Q_{\alpha }$,
\begin{equation}
\begin{array}{ll}
\delta _{\varepsilon }e^{a}=\frac{1}{2}\bar{\epsilon}\Gamma ^{a}\psi , &
\delta _{\varepsilon }\psi =D\epsilon, \;\delta _{\varepsilon
}\omega ^{ab}=0,\; \\
\delta _{\varepsilon }b^{ab}=\frac{1}{2}\bar{\epsilon}\Gamma ^{ab}\psi , &
\delta _{\varepsilon }b^{abcde}=\frac{1}{2}\bar{\epsilon}\Gamma ^{abcde}\psi
.
\end{array}
\label{susytransf}
\end{equation}
Thus, the supersymmetric extension of (\ref{I_G}), invariant under (\ref
{susytransf}) is found to be
\begin{eqnarray}
I_{\alpha } &=&I_{G}+I_{\psi }-\frac{\alpha }{6}%
\int_{M_{11}}R_{abc}R_{de}b^{abcde}  \nonumber \\
&&+8(1-\alpha )\int_{M_{11}}[R^{2}R_{ab}-6(R^{3})_{ab}]R_{cd}\left( \bar{\psi%
}\Gamma ^{abcd}D\psi -12R^{[ab}b^{cd]}\right) ,  \label{Action}
\end{eqnarray}
where $R^{2}:=R^{ab}R_{ba}$ and $(R^{3})^{ab}:=R^{ac}R_{cd}R^{db}$. Here $%
\alpha $ is a dimensionless constant whose meaning will be discussed below.

This action is invariant under (\ref{PoincareTrans}), (\ref{susytransf}),
local Lorentz rotations, and also under the local Abelian transformations
\begin{equation}
\begin{array}{ll}
\delta b^{ab}=D\theta ^{ab}, & \delta b^{abcde}=D\theta ^{abcde}
\end{array}
.  \label{ZZ}
\end{equation}
Invariance under general coordinate transformations is guaranteed by the use
of forms. It is simple to see that the local invariances of the action,
including Poincar\'{e} transformations, supersymmetry (\ref{susytransf})
together with (\ref{ZZ}), are a gauge transformation for the M-connection (%
\ref{M-connection}) with parameter $\lambda =\lambda ^{a}P_{a}+\frac{1}{2}%
\lambda ^{ab}J_{ab}+\theta ^{ab}Z_{ab}+\theta ^{abcde}Z_{abcde}+1/\sqrt{2}%
\epsilon ^{\alpha }Q_{\alpha }$. As a consequence, the invariance of the
action under the supersymmetry algebra is ensured by construction without
invoking field equations or requiring auxiliary fields.


\subsection{Manifest M-Covariance}

The action (\ref{Action}) describes a gauge theory for the M-algebra with
fiber bundle structure, which can be seen explicitly by writing the
Lagrangian as a Chern-Simons form \cite{FootnoteCS} for the M-connection (%
\ref{M-connection}). Indeed, the Lagrangian satisfies $dL=\left\langle
F^{6}\right\rangle $, where the curvature $F=dA+A^{2}$ is given by
\[
F=\frac{1}{2}R^{ab}J_{ab}+\tilde{T}^{a}P_{a}+1/\sqrt{2}D\psi ^{\alpha
}Q_{\alpha }+\tilde{F}^{^{[2]}}Z_{^{[2]}}+\tilde{F}^{^{[5]}}Z_{^{[5]}},
\]
with $\tilde{T}^{a}=De^{a}-(1/4)\bar{\psi}\Gamma ^{a}\psi $ and $\tilde{F}%
^{[k]}=Db^{[k]}-(1/4)\bar{\psi}\Gamma ^{[k]}\psi $ for$\,\,k=2$, $5$. The
bracket $\left\langle ...\right\rangle $ stands for a multilinear form of
the M-algebra generators $G_{A}$ whose only nonvanishing components are
given by
\[
\begin{array}{l}
\left\langle J_{a_{1}a_{2}},\cdots ,J_{a_{9}a_{10}},P_{a_{11}}\right\rangle =%
\frac{16}{3}\epsilon _{a_{1}\cdots a_{11}}\;, \\
\left\langle J_{a_{1}a_{2}},\cdots ,J_{a_{9}a_{10}},Z_{abcde}\right\rangle
=-\alpha \frac{4}{9}\epsilon _{a_{1}\cdots a_{8}abc}\eta
_{[a_{9}a_{10}][de]}\;, \\
\left\langle
J_{a_{1}a_{2}},J_{a_{3}a_{4}},J_{a_{5}a_{6}},J^{a_{7}a_{8}},J^{a_{9}a_{10}},Z^{ab}\right\rangle =(1-\alpha )%
\frac{16}{3}\left[ \delta _{a_{1}\cdots \cdots a_{6}}^{a_{7}\cdots
a_{10}ab}-\delta _{a_{1}\cdots a_{4}}^{a_{9}a_{10}ab}\delta
_{a_{5}a_{6}}^{a_{7}a_{8}}\right] \\
\left\langle
Q,J_{a_{1}a_{2}},J^{a_{3}a_{4}},J^{a_{5}a_{6}},J^{a_{7}a_{8}},Q\right\rangle
=\frac{32}{15}\left[ C\Gamma _{a_{1}a_{2}}^{\;\,\,\quad
\;\,a_{3}\cdots a_{8}}+\right. \\
\;\;\;\;\;\;\;\;\;\;\;\;\;\;\;\;\;\;\;\;\;\;\;\;\;\;\;\;\;\;\;\;\;\;\;\;\;\;%
\;\;\;\qquad \;\;\;\;\;\;\;\;\left. (1-\alpha )\left( 3\delta
_{a_{1}a_{2}ab}^{a_{3}\cdots a_{6}}C\Gamma ^{a_{7}a_{8}ab}+2C\Gamma
^{a_{3}\cdots a_{6}}\delta _{a_{1}a_{2}}^{a_{7}a_{8}}\right) \right] ,
\end{array}
\]
where (anti-)symmetrization under permutations of each pair of generators is
understood when all the indices are lowered. The existence of this bracket
allows writing the field equations in a manifestly covariant form as
\begin{equation}
\left\langle F^{5}G_{A}\right\rangle =0.  \label{FieldEqs}
\end{equation}
In addition, if the eleven-dimensional spacetime is the boundary of a
twelve-dimensional manifold, $\partial \Omega _{12}=M_{11}$, the action (\ref
{Action}) can also be\ written as $I=\int_{\Omega _{12}}\left\langle
F^{6}\right\rangle $, which describes a topological theory in twelve
dimensions. In spite of its topological origin, the action does possess
propagating degrees of freedom and hence it should not be thought of as a
topological field theory.


\section{Gravitons and four-dimensional spacetime}

We now turn to the problem of identifying the true vacuum of the theory.
Obviously, a configuration with a locally flat connection, $F=0$, solves the
field equations\ and would be a natural candidate for vacuum in a standard
field theory. However, no local degrees of freedom can propagate on such
background because all perturbations around it are zero modes. Note that
eleven-dimensional Minkowski spacetime is maximally supersymmetric by virtue
of (\ref{susytransf}), however as it obeys $F=0$, the propagators on it are
ill-defined, and hence it is a sort of false vacuum.

In a matter-free configuration, Eq. (\ref{FieldEqs}) is a set of quintic
polynomials for the Riemann two-form $R^{ab}$. The dynamical field equations
take the form
\begin{eqnarray}
\epsilon^{0ij_1\cdots j_9}\left\langle F_{j_1j_2}\cdots F_{j_7j_8}
\left(\partial_t A_{j_9}-\nabla_{j_9}A_0 \right) G_A\right\rangle =0
\end{eqnarray}
So, in order to have propagation for $A_{j}$, the spatial components $F_{ij}$
cannot be small. Hence, a deviation around $F=0$ that propagates cannot be
infinitesimal and is therefore non-perturbative and non-local. A necessary
condition to have well-defined perturbations is that the background solution
be a simple zero of at least one of the polynomials. In particular, this
requires the curvature to be nonvanishing on a submanifold of a large enough
dimension.

Let us consider a torsionless spacetime with a product geometry of the form $%
X_{d+1}\times S^{10-d}$, where $X_{d+1}$ is a domain wall whose worldsheet
is a $d$-dimensional constant curvature spacetime $M_{d}$. The line element
is given by

\begin{equation}
ds^{2}=\exp \left( -2a|z|\right) \left( dz^{2}+\tilde{g}_{\mu \nu
}^{(d)}(x)dx^{\mu }dx^{\nu }\right) +\gamma _{mn}^{(10-d)}(y)dy^{m}dy^{n},
\label{Ansatz}
\end{equation}
where $\tilde{g}_{\mu \nu }^{(d)}$ stands for the worldsheet metric with $%
\mu ,\nu =0,...,d-1$; $\gamma _{mn}^{(10-d)}$ is the metric of $S^{10-d}$ of
radius $r_{0}$ and $a$ is a constant.

This Ansatz solves the vacuum field equations provided the projection of the
Riemann tensor along the worldsheet,
\[
R^{ij}=\tilde{R}^{ij}-a^{2}\,\tilde{e}^{i}\wedge \tilde{e}^{j},
\]
vanishes (here $\tilde{e}^{i}$ and $\tilde{R}^{ij}$ stand for the vielbein
and the Riemann curvature of the worldsheet, respectively). This means that $%
M_{d}$ is either locally de Sitter spacetime of radius $a^{-1}$, or locally
Minkowski for $a=0$.

The requirement that the curvature of (\ref{Ansatz}) be a simple zero,
implies, after a straightforward computation, that $d$ cannot be greater
than four. Then, the condition of having well-defined propagators singles
out the dimension of the worldsheet to be $d=4$, and $a^{2}>0$. Indeed, for $%
d=4$, the only relevant equation for the perturbations is the one that
arises from the variation with respect to $\tilde{e}^{i}$,
\begin{equation}
a\delta (z)\epsilon _{ijkl}\delta (\tilde{R}^{jk}-a^{2}\,\tilde{e}^{j}\tilde{%
e}^{k})\tilde{e}^{l}=0.  \label{GB}
\end{equation}
Since for $a=0$ this equation becomes empty, Minkowski spacetime must be
ruled out. Thus, the existence of the propagator requires the
four-dimensional cosmological constant to be strictly positive and given by $%
\Lambda _{4}=3a^{2}$.

Note that Eq. (\ref{GB}) has support only on the $z=0$ plane. Perturbations
along the worldsheet, $\delta \tilde{g}_{\mu \nu }=h_{\mu \nu }(x)$
reproduce the linearized Einstein equations in four-dimensional de Sitter
spacetime. The modes that depend on the coordinates transverse to the
worldsheet fall into two classes. Those of the form $\delta \tilde{g}_{\mu
\nu }=h_{\mu \nu }(x,y)$ are massive Kaluza-Klein modes with a discrete
spectrum, while $\delta \tilde{g}_{\mu \nu }=h_{\mu \nu }(x,z)$ correspond
to Randall-Sundrum-like massive modes whose spectrum is continuous and has a
mass gap. The perturbations of the remaining metric components are zero
modes, which is related to the fact that the equations are not deterministic
for the compact space. A detailed analysis of this, as well as of the
perturbations of matter fields will be presented elsewhere \cite{HTZ}.


\section{Discussion}


We have presented a framework in which the spacetime dimension is
dynamically selected to be four. The\ mechanism is based on a new
eleven-dimensional action of the Chern-Simons type, which is a gauge theory
for the M-algebra. The possibility of dynamical dimensional reduction arises
because the theory has radically different spectra around backgrounds of
different effective spacetime dimensions. Thus, in a family of product
spaces of the form $X_{d+1}\mathbf{\times }S^{10-d}$, the only option that
yields a well defined low energy propagator for the graviton is $d=4$ and $%
\Lambda _{4}>0$. It should be stressed that for all gravity theories of the
type discussed here, possessing local Poincar\'{e} invariance in dimensions $%
D=2n+1\geq 5$, four-dimensional de Sitter spacetime is also uniquely
selected by the same mechanism as the background for the low energy
effective theory.

The action discussed in this paper has a free parameter $\alpha $, which
reflects the fact that the theory contains two natural limits which
correspond to different subalgebras of (\ref{MAlgebra}). For $\alpha =0$,
the action $I_{0}$ in Eq. (\ref{Action}) does not depend on $b^{[5]}$ and
corresponds to a gauge theory for the supermembrane algebra, while for $%
\alpha =1$, the bosonic field $b^{[2]}$ decouples, and $I_{1}$ is a gauge
theory for the super five-brane algebra as discussed in \cite{BTZ}. It is
interesting to note that the linear combination of both limits, $I_{\alpha
}=I_{0}+\alpha (I_{1}-I_{0})$, is not only invariant under the intersection
of both algebras, but under the entire M-algebra. As the term $I_{1}-I_{0}$
does not couple to the vielbein and is invariant under supersymmetry by
itself, $\alpha $ is an independent coupling constant. A similar situation
occurs in nine dimensions where, in one limit, the theory corresponds to the
super five-brane algebra, while for the other it is a gauge theory for the
super-Poincar\'{e} algebra with a central extension \cite{HOT}.

In the presence of negative cosmological constant, the eleven-dimensional
AdS supergravity presented in Ref. \cite{Tr-Z} can be written as a
Chern-Simons theory for $osp(32|1)$, which is the supersymmetric extension
of AdS$_{11}$. It is natural to ask whether there is a link between that
theory in the vanishing cosmological constant limit, and the one discussed
here. Since the M-algebra has $55$ bosonic generators more than $osp(32|1)$,
these theories cannot be related through a In\"{o}n\"{u}-Wigner contraction
for a generic value of $\alpha $. However, it has been recently pointed out
in \cite{AIPV}, generalizing the procedure of \cite{Hatsuda-Sakaguchi}, that
it is possible to obtain the M-algebra from an expansion of $osp(32|1)$. In
this light, applying this procedure to the eleven-dimensional AdS
supergravity theory, it should be expected that the action presented here
will be recovered up to some additional terms decoupled from the vielbein,
that are supersymmetric by themselves.

\emph{Acknowledgments.-} We thank J. Edelstein, G. Kofinas, C. Mart\'{i}nez,
C. N\'{u}\~{n}ez and C. Teitelboim for many enlightening discussions. This
work is partially supported by grants 3020032, 1010450, 1010446, 1010449,
1020629 and 1040921 from FONDECYT. Institutional support to the Centro de
Estudios Cient\'{i}ficos (CECS) from Empresas CMPC is gratefully
acknowledged. CECS is a Millennium Science Institute and is funded in part
by grants from Fundaci\'{o}n Andes and the Tinker Foundation.


\begin{thebibliography}{99}
\bibitem{Footnote1}  Under the transformations (\ref{PoincareTrans}), the
Einstein-Hilbert action $I_{EH}=\int \epsilon _{a_{1}\cdots
a_{11}}R^{a_{1}a_{2}}e^{a_{3}}\cdots e^{a_{11}}$ changes by a term
proportional to $\int \epsilon _{a_{1}\cdots
a_{11}}R^{a_{1}a_{2}}T^{a_{3}}e^{a_{4}}\cdots e^{a_{10}}\lambda ^{a_{11}}$,
which vanishes only if the torsion $T^{a}=De^{a}$ does. However, this last
condition is incompatible with the transformations because $\omega ^{ab}$
would be a function of $e^{a}$ and hence, its variation could not vanish.
See T. Regge, Phys. Rept. \textbf{137}, 31 (1986).
\end{thebibliography}
\end{document}